%% file: main.tex
\documentclass[acmsmall,screen]{acmart}

\setcopyright{acmlicensed}
\copyrightyear{2024}

\acmYear{2024}
\acmJournal{TOSEM}
\acmDOI{XXXXXXX.XXXXXXX}\renewcommand\footnotetextcopyrightpermission[1]{} 

\AtBeginDocument{%
  \providecommand\BibTeX{{%
    \normalfont B\kern-0.5em{\scshape i\kern-0.25em b}\kern-0.8em\TeX}}}

\usepackage{tcolorbox}
\newenvironment{limitation}[1]{
	\begin{tcolorbox}[colback=yellow!10!white, 
  colframe=yellow!10!white, 
 arc=0mm,grow to left by=0mm,left=0mm,grow to right by=0mm,left=0mm,right=0mm,top=0mm,bottom=0mm]
		\textbf{Limitation #1:}
	}
	{
	\end{tcolorbox}
}
\usepackage{subcaption} 
\usepackage{graphicx}
\definecolor{gray50}{gray}{.5}
\definecolor{gray40}{gray}{.6}
\definecolor{gray30}{gray}{.7}
\definecolor{gray20}{gray}{.8}
\definecolor{gray10}{gray}{.9}
\definecolor{gray05}{gray}{.95}

\newlength\Linewidth
\def\findlength{\setlength\Linewidth\linewidth
	\addtolength\Linewidth{-4\fboxrule}
	\addtolength\Linewidth{-3\fboxsep}
}

 \newenvironment{solution}[1]{
	\begin{tcolorbox}[colback=green!10!white, 
  colframe=green!10!white, 
 arc=0mm,grow to left by=0mm,left=0mm,grow to right by=0mm,left=0mm,right=0mm,top=0mm,bottom=0mm]
		\textbf{Solution #1:}
	}
	{
	\end{tcolorbox}
}

\usepackage{xspace}
\usepackage[utf8]{inputenc}
\usepackage{multirow}
\usepackage{tabularx}

\newcommand{\etal}{\emph{et~al.}\xspace}
\newcommand{\eg}{\emph{e.g.,}\xspace}

\newcommand{\AI}{\emph{HyperAssistant}\xspace}

\begin{document}

\title[From Today's Code to Tomorrow's Symphony: The AI Transformation of Developer's Routine by 2030]{From Today's Code to Tomorrow's Symphony:\\The AI Transformation of Developer's Routine by 2030}


\author{Ketai Qiu}
\affiliation{%
  \institution{Università della Svizzera Italiana (USI)}
  \city{Lugano}
  \country{Switzerland}}
\email{ketai.qiu@usi.ch}

\author{Niccolò Puccinelli}
\affiliation{%
  \institution{Università della Svizzera Italiana (USI)}
  \city{Lugano}
  \country{Switzerland}}
\email{niccolo.puccinelli@usi.ch}

\author{Matteo Ciniselli}
\affiliation{%
  \institution{Università della Svizzera Italiana (USI)}
  \city{Lugano}
  \country{Switzerland}}
\email{matteo.ciniselli@usi.ch}

\author{Luca Di Grazia}
\affiliation{%
  \institution{Università della Svizzera Italiana (USI)}
  \city{Lugano}
  \country{Switzerland}}
\email{work@lucadigrazia.com}

\renewcommand{\shortauthors}{Qiu, et al.}

\begin{abstract}
In the rapidly evolving landscape of software engineering, the integration of Artificial Intelligence (AI) into the Software Development Life-Cycle (SDLC) heralds a transformative era for developers. Recently, we have assisted to a pivotal shift towards AI-assisted programming, exemplified by tools like GitHub Copilot and OpenAI’s ChatGPT, which have become a crucial element for coding, debugging, and software design. In this paper we provide a comparative analysis between the current state of AI-assisted programming in 2024 and our projections for 2030, by exploring how AI advancements are set to enhance the implementation phase, fundamentally altering developers' roles from manual coders to orchestrators of AI-driven development ecosystems. We envision \textit{HyperAssistant}, an augmented AI tool that offers comprehensive support to 2030 developers, addressing current limitations in mental health support, fault detection, code optimization, team interaction, and skill development. We emphasize AI as a complementary force, augmenting developers' capabilities rather than replacing them, leading to the creation of sophisticated, reliable, and secure software solutions. Our vision seeks to anticipate the evolution of programming practices, challenges, and future directions, shaping a new paradigm where developers and AI collaborate more closely, promising a significant leap in SE efficiency, security and creativity.
\end{abstract}

\ccsdesc[500]{Software and its engineering~Automatic programming}

\keywords{Software Engineering, AI for Code, Human Factors in Software Engineering}


\maketitle

\input{sections/introduction}
\input{sections/dev2024}

\input{sections/dev2030}
\input{sections/2024vs2030}
\input{sections/conclusion}

\bibliographystyle{ACM-Reference-Format}
\bibliography{main}

\end{document}

%% file: sections/introduction.tex
\section{Introduction}

\begin{figure}[t]
	\centering
	\includegraphics[width=.5\linewidth]{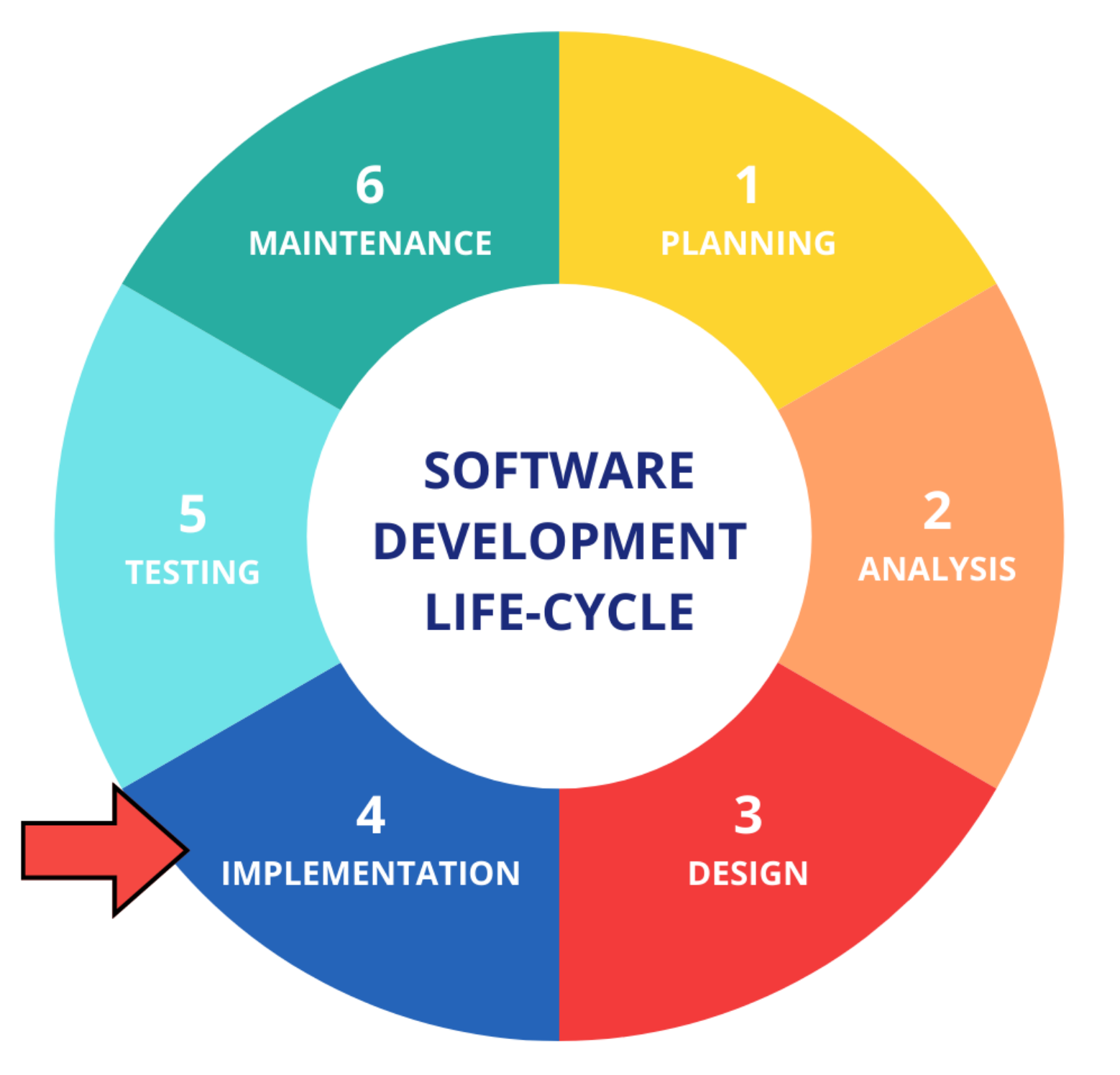}
	\caption{The Software Development Life Cycle. We specifically focus on the implementation phase.}
	\label{fig:software-cycle}
\end{figure}

\paragraph{Context.} 
The evolution of software engineering and the integration of AI assistants, like GitHub Copilot~\cite{chen:arxiv2021} and ChatGPT~\cite{openai2024gpt4}, is dramatically changing daily routines of software developers~\cite{Barenkamp:AIP:2020,Latinovic:2021}.
Several studies investigated the usage of these tools in the Software Development Life-Cycle (SDLC) (Figure ~\ref{fig:software-cycle}), evaluating how developers leverage them, showing the unprecedented support in coding, debugging, and even in the creative aspects of software design~\cite{Moradi:JSS:2023,Sridhara:arXiv:2023,Peng:arXiv:2023}.

This change in the SDLC is evident across all stages, from planning to maintenance, and promises to enhance the overall software quality. For example, at the implementation stage, tools like GitHub Copilot and OpenAI's ChatGPT already offer real-time coding assistance and suggest optimizations, reducing the development time and improving code quality. AI advancements have also revolutionized the testing phase of the SDLC~\cite{Wang2024, Dohmke:arXiv:2023}. AI can automatically generate test cases based on the requirements~\cite{Mustafa:CMC:2021} and code~\cite{Shin:arXiv:2024}, ensuring comprehensive coverage and identifying edge cases that might be overlooked by human testers. Finally, integrating AI into the SDLC provides benefits in the continuous learning mechanism. AI tools can learn from each project, continuously improving their suggestions and assistance, and consequently the overall software development.

\paragraph{Significance.}
The significance of leveraging AI in software engineering embodies substantial cost savings and innovation in a competitive market, by enhancing efficiency, reducing the incidence of bugs and accelerating the development time. These factors and the recent huge investments into this domain underscore the potential and the importance of AI in reshaping the future of software development~\cite{Riberio2023}.
Our vision is rooted in the belief that AI, much like a co-pilot in the cockpit, serves as an indispensable assistant rather than a replacement for human developers. By augmenting the capabilities of software engineers through AI, we anticipate a future where the synergy between human and AI will achieve the creation of sophisticated, reliable, and more secure software solutions.

\paragraph{Contributions.} 
The importance of focusing on the implementation (coding) phase cannot be overstated. This phase is critical in determining the quality, reliability, and functionality of software products. Our paper contributes to the ongoing discourse on the integration of AI in the SDLC~\cite{Sorte:IJTM:2015} with a particular focus on this pivotal phase. Through a comparative analysis between current practices (2024) and future projections (2030), we aim to highlight the impact of AI tools on the implementation stage of the SDLC. This analysis will provide insights into how these advancements not only streamline processes but also fundamentally change the role of developers, shifting from manual coders to orchestrators of AI-driven development ecosystems. By examining the evolution of the implementation phase, we shape a future where developers and AI collaborate more closely, marking a significant leap forward in the field of SE. In summary, our main contributions are:
\begin{itemize}
    \item A comparative analysis to evaluate the impact of AI on the implementation phase of the SDLC, contrasting current practices in 2024 with projections for 2030.
    \item We discuss how advancements in AI not only streamline development processes but also significantly alter the role of developers, shifting them from manual coding to orchestrating AI-driven ecosystems.
    \item We envision a future where the integration of AI fosters a closer collaboration between developers and technology in the SDLC, representing a pivotal advancement in software engineering.
\end{itemize}

\paragraph{Paper Structure.}
The paper is organized as follows. Section~\ref{sec:2024only} provides an historical introduction about programming, overviewing the limitations of the current approaches. Section~\ref{sec:2030only} envisions how current limitations could be overcome. thanks to the introduction of \AI, an innovative augmented assistant that, by 2030, will possess the capability to provide comprehensive support to developers. Section~\ref{sec:2024vs2030} describes a hypothetical workday for a developer in 2024 and 2030, focusing on the limitations of the AI in 2024. Section~\ref{sec:discussion} discusses the implications for developers and research in the upcoming years. Finally, Section~\ref{sec:conclusion} summarizes the content of the paper and concludes the work.

%% file: sections/dev2024.tex
\section{Developers in 2024}
\label{sec:2024only}
The integration of AI with software engineering has unlocked new pathways for tackling a broad spectrum of challenges across various levels of abstraction. Concurrent with the advancements in hardware, particularly in Graphics Processing Units (GPUs)~\cite{Ang:Electronics:2021}, and the progression of techniques ranging from Machine Learning (ML) and Deep Learning (DL)~\cite{Zhang:JIII:2021} to Large Language Models (LLMs)~\cite{Li:ACE:2024}, there has been a significant increase in efforts to leverage AI in software development. This synergy not only enhances the efficiency and effectiveness of software solutions but also paves the way for innovative approaches to complex problem-solving in software engineering.
\subsection{Historical Evolution of Programming}

\input{sections/dev2024_history}

\subsection{Related Work}
\input{sections/dev2024_related_work}

\subsection{Limitations of the Current Approaches}
\input{sections/dev2024_today}

%% file: sections/dev2024_history.tex
Retrospecting the programming history, we can identify three distinct ages during the evolution of programming: pre-APIs, APIs and LLMs, which witness the transition of software development from monolithic to microservice-based, and then to intellectual applications that involve enormous natural interactions between humans and software.

Before the advent of APIs, software applications were designed to operate in a standalone way, which means they were not able to interact with other systems. In other words, the interactions among different components of the system are tightly coupled with specific workflows during the era of pre-APIs. Although it is convenient for developers to build a monolithic application at the beginning of the software development, the consequential product will severely suffer from the lack of extensibility and flexibility.

Therefore, APIs are proposed to overcome limitations of monolithic applications~\cite{Sousa2006}. APIs define a collection of protocols for the standardized communication between related systems. Since different systems are very likely implemented using different frameworks or programming languages, it is important to have a consistent and controllable way to share information with each other. Currently, REST (Representational State Transfer) and GraphQL\footnote{\href{https://graphql.org/}{https://graphql.org/}} are the most frequently used standards to design APIs for production usage. APIs are independent from the implementations of the interoperated systems, so they are flexible and scalable~\cite{Blinowski2022}. These characteristics also accelerate for software companies the transition of software development from monolithic to microservice-based architecture, where the interactions between distinct microservices are crucial~\cite{AUER2021106600}.

Recently, LLMs are becoming a new manner of programming due to their incredible performance and convenience. LLMs are a special type of GAI system that is capable of generating texts based on next token prediction. Hence, they're specially suitable for programming tasks with hundreds of lines of code. They're designed to be efficient, scalable, and flexible. These 3 unique advantages make LLMs highly popular for programming nowadays.

Firstly, LLMs are efficient.
With the help of LLMs, anyone can develop software applications even without domain knowledge by simply communicating with AI chatbots. End users can draft a method or a class in a few minutes by explaining, using the natural language, their requirements to LLMs, that are able to translate them into the implementation~\cite{kazemitabaar2023}. 
This communication process is officially called Prompt Engineering. From this perspective, it's essential to know how to efficiently convey specific requirements with LLMs. There are mainly two categories of prompts: zero-shot and few-shot, where ``shot'' refers to the simple explanatory example (input and expected output) within the relevant context provided by the users. With zero-shot learning, users do not provide any example to the model, relying on the former knowledge acquired by the model during the training, while in the few-shot learning, the model is provided with some clarifying examples.

Secondly, LLMs are scalable. Considering the ease of use of LLMs, it is essential to deploy LLMs on the cloud in a distributed way to serve requests from a handful of programmers. This can be achieved via distributing model shards across multiple GPUs~\cite{lin2024efficientllmtrainingserving} and caching prompts and the corresponding generated tokens as key-value pairs~\cite{NEURIPS2023LiuLLMCache}. However, some developers prefer to deploy LLMs on local devices to secure confidential information during inference. Quantization is designed to facilitate this process by reducing the precision of the model's weights. Since LLMs usually have billions of parameters, users can save a significant amount of memory by using 8-bit floats or even 4-bit floats instead of 16-bit floats to store weights. There are several techniques (e.g., ZeroQuant~\cite{Yao:ZeroQuant:NEURIPS:2022} and GPTQ~\cite{Elias:GPTQ:2022}) proposed to quantize LLMs while maintaining the performance.

Thirdly, LLMs are flexible. Since LLMs are trained with a large volume of data collected from the Internet with a general purpose of inference, they may not perform well for a specific programming language. But they can be further tailored via fine-tuning or retrieval-augmented generation (RAG). For example, CodeLlama-70B-Python is built on top of Llama 2 for Python-related tasks\footnote{\url{https://ai.meta.com/blog/code-llama-large-language-model-coding/}}, which is an ideal tool for Python developers. Similarly, developers using other languages can easily adapt available LLMs for their specific programming tasks with few fine-tuning efforts.

In a nutshell, LLMs can not only guide junior programmers via tutorial conversations but also help senior developers accelerate the code understanding and the development process in an even faster agile manner, because they can delegate tedious coding parts to LLMs and mainly focus on the critical business logic~\cite{Nam2024}.

%% file: sections/dev2024_related_work.tex
Back in 2012, Ammar \etal~\cite{Ammar:2012} surveyed the integration of AI techniques into software engineering processes, aiming to reduce development time and improve software quality. By seeking to bridge the gap between research and practice in applying AI to software engineering, they focused on requirements analysis, architecture design, coding, and testing, highlighting practical challenges and recent research in the area. Concurrently, Harman~\cite{Harman:RAISE:2012} directed attention towards a heightened level of abstraction, emphasizing the evolution within software engineering from conventional, localized, and clearly delineated construction methods towards the orchestration of expansive, interconnected, and intelligent systems. Harman’s perspective focused on the several challenges ahead for AI integration in software engineering, such as the ways in which AI techniques can be used to gain insight to software engineers and the need for balancing automation with human intervention.

Three years later, Sorte \etal~\cite{Sorte:IJTM:2015} provided a comprehensive overview of how AI techniques are integrated into various phases of the Software Development Life Cycle (SDLC) to automate and enhance the process. The authors explored the intersection of AI and software engineering, revealing that despite their separate development, these fields have much to offer each other. The paper identifies key areas where AI contributes to software engineering (e.g., requirement specification, design, code generation, testing), while discussing relevant specific AI techniques for each phase of the SDLC. More recently, Shehab \etal~\cite{Shehab:IJCIM:2020} outlined the integration of AI into the SDLC, underscoring the potential of ML to enhance various phases of the software engineering process, including requirements engineering and code generation.

Nevertheless, in recent times, we have observed yet another paradigm shift with the emergence of Generative AI (GAI), which promises to significantly increase productivity~\cite{Noy:Science:2023} through the exploitation of natural language. Leading the charge are tools like OpenAI's ChatGPT, Github Copilot, and Google Gemini, which have become fundamental for developers owing to their remarkable capacity to augment productivity, foster creativity, and streamline efficiency~\cite{Ebert:IEEE:2023, Peng:arXiv:2023}. In order to understand how programmers interact with these system, Mozannar \etal~\cite{Mozannar:arXiv:2022} introduced the CodeRec User Programming States (CUPS) taxonomy, aimed at categorizing prevalent activities undertaken by programmers when utilizing these AI tools. The study involved 21 programmers who completed coding tasks and retrospectively labeled their sessions with CUPS categories. Key findings revealed significant time allocations towards activities tailored to interacting with Copilot. Notably, programmers frequently deferred suggestion verification, resulting in a notable portion of session time dedicated to managing Copilot's suggestions. These insights shed light on the inefficiencies and time overheads associated with the utilization of such systems. 

In the realm of developer support, AI assistance has become a focal point of discussion. The rapid advancement of GAI in recent years prompts speculation on the extent of its future development and its potential to address researchers' concerns. Simultaneously, people are increasingly embracing and adapting these tools, as evidenced by the emergence of prompt engineering~\cite{Denny:SIGCSE:2023}, which is gradually narrowing the divide between software and human interaction.

%% file: sections/dev2024_today.tex

The advent of generative AI as a developer's assistant has also introduced several challenges and opportunities among researchers and practitioners. 

In the world of software development as in any other job, maintaining optimal mental health is crucial for sustained productivity, creativity, and overall happiness~\cite{Graziotin2018}. As developers navigate through intricate codebases and tight deadlines, the demands of the job can often take a toll on their mental and physical health. However, at this stage AI assistants focus mostly on solving technical challenges than human-factor problems related to coding activities.


\begin{limitation}
    1~The mental health of programmers is frequently overlooked, despite its significant impact on productivity and overall well-being. 
\end{limitation}

Despite the significant potential of AI in enhancing the software engineering process, especially with the recent advancements of automated code generation~\cite{Odeh:TEM:2024}, over-reliance on such technology poses a risk in terms of security and code vulnerability. For example, the analysis of Pearce \etal~\cite{Pearce:IEEE:2022} revealed that approximately 40\% of the generated programs with GitHub Copilot contains vulnerabilities and researchers are working on innovative solutions to avoid security issues introduced by AI generated code~\cite{Vechev2023}. Perry \etal~\cite{Perry:ACM:2023} investigated whether developers relying on AI code assistants, like GitHub Copilot, produce less secure code than those who do not. By conducting a user study involving 47 participants, they found that those with access to an AI assistant produced significantly less secure solutions compared to those without access. On the other hand, the paper of Asare \etal~\cite{Asare:ESE:2023} revealed that Copilot's likelihood of introducing vulnerabilities varies with the type of vulnerability and, most importantly, that GAI for code generation, while not perfect, does not perform worse than human developers in introducing vulnerabilities. These insights unquestionably establish a groundwork for further research in this domain.


\begin{limitation}
    2~Developers often overestimate the capabilities of tools like GitHub Copilot by deferring the verification of the generated code, introducing more vulnerabilities and bugs. 
\end{limitation}

Another relevant aspect to consider is the limits of these models in understanding semantic information. 
For example, Nie \etal~\cite{nie:icse2023} showed the significant improvement of the quality of the code generated for software testing when providing additional semantic information, like similar statements or the types of the variable defined, that can easily inferred by a developer but not by AI tools.
Moreover, developers did not attribute the right importance to the code optmization, often using the copy and paste mechanism while programming.
Several studies investigate this phenomenon \cite{kim:isese2004,li:tse2006,kapser:ese2008,baker:wcre1995}. Kim \etal~\cite{kim:isese2004} showed that, despite the copy mostly involve single statements, snippets are copied in 25\% of the cases, while Baker~\cite{baker:wcre1995} reported the seek of performance as reason behind that, with developers that are evaluated on their productivity, pushing them to verbatim copy code rather than promoting refactoring of old code. Sometimes this behavior is unintention, with developers that tend to re-implement the same code since they are not aware of the presence of the the same snippet~\cite{kapser:ese2008}.

This is reflected on the limited support for challenging tasks of the software development, like the optimization or the refactoring of the code.


\begin{limitation}
    3~Modern tools frequently face challenges in code optimization, resulting in misleading errors and misunderstandings when developers fail to supply the necessary context.
\end{limitation}

Development's projects involve several developers, each of one contributing based on their own skills. For example, the developer that is most familiar with the database management is in charge of building a reliable data storing infrastructure, while the one with a lot of experience in the front end will develop an engaging web page. Dividing these activities among different developers is far from trivial, and AI assistants are not able to improve the interaction between the team members, for example scheduling a meeting when a developer is struggling with the task at hand. 

\begin{limitation}
    4~AI tools are not helpful in promoting a synergical interaction between the team member, thus boosting the overall performance of the team.
\end{limitation}

Finally, the AI models are not fully integrated in the life of developers, and they are not able to take informed decision based on each developer, for example favoring new personalized learning paths.
Programming languages are evolving over time. For example, in the last 30 years, more than 20 different Java versions have been released. Each version introduces new features (\eg the \emph{lambda expression} for Java 8) and makes obsolete some of them.
AI tools are mostly trained on past data, and usually recommend popular solutions, lacking the capability to harness recent language advancements and present an innovative solution to the problem.



\begin{limitation}
    5~Actual models fail to consider the unique needs and skills of programmers, lacking on personalized learning resources for software engineers.
\end{limitation}

%% file: sections/dev2030.tex
\section{Developers in 2030}
\label{sec:2030only}

\begin{figure}[t]
	\centering
	\includegraphics[width=.75\linewidth]{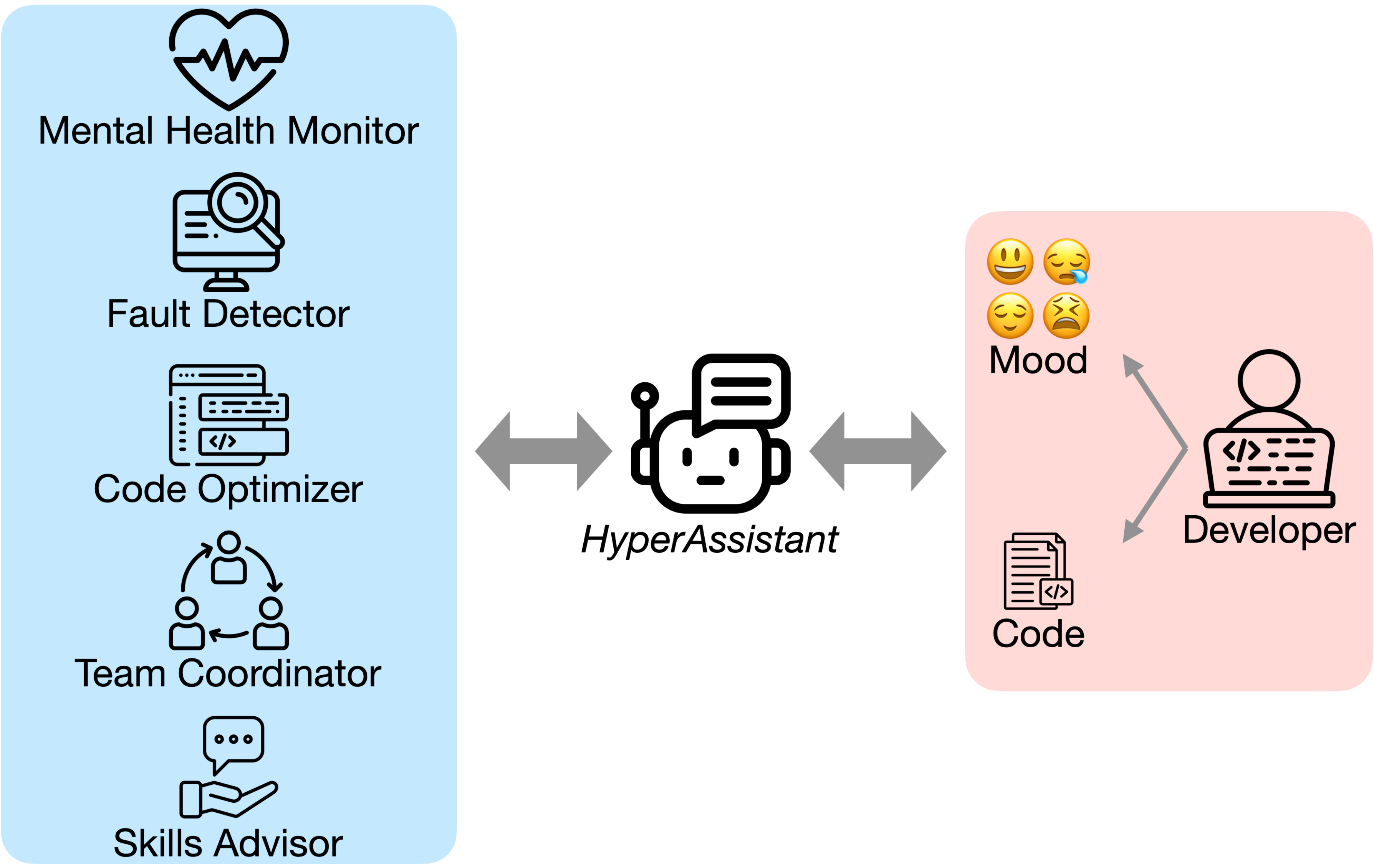}
	\caption{Overview of \AI workflow and its components to improve developer productivity.}
	\label{fig:hyperassistant}
\end{figure}

In the previous section, we present the limitations of the novel tools in the Software Engineering tasks. Despite the good results achieved, they can still enhance the support provided to developers during the daily routine. In this section, we envision how developers may benefit from \AI, a hypothetical augmented assistant able to fully support developers in 2030. The complete structure of \AI is shown in Figure \ref{fig:hyperassistant}.

Our proposed system, i.e., \AI,  is composed of five subsystems. 
The Mental Health Monitor is specifically designed for tracking the mental state of developers. 
The Fault Detector is used to guarantee quality and reliability of the software under development.
The Code Optimizer aims to accelerate the coding phase via automatic code completion and automatic code review.
The Team Coordinator helps reduce invalid or repetitive communication as much as possible.
The Skills Advisor facilitates continuous learning for developers.
Each subsystem takes the developer's current mood and code into account and generates suggestions accordingly like a chat agent.
The functionality and the underlying idea of each subsystem is described in detail as follows.

\begin{figure*}[t]
	\centering
	\begin{subfigure}[t]{.45\linewidth}
		\includegraphics[width=\linewidth]{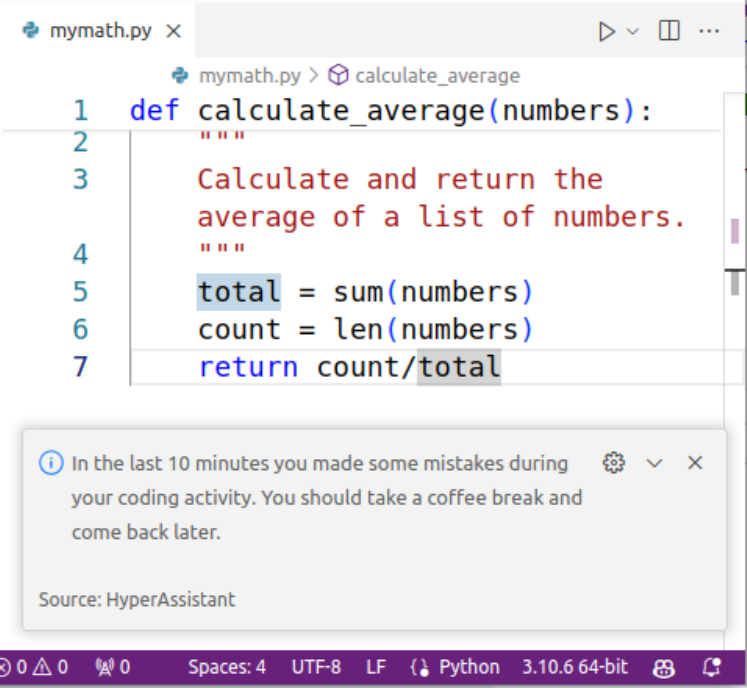}
		\caption{AI suggests that the developer take a break after it detects signs of tiredness.}
		\label{fig:mentalhealth}
	\end{subfigure}
	\hfill 
	\begin{subfigure}[t]{.46\linewidth}
		\includegraphics[width=\linewidth]{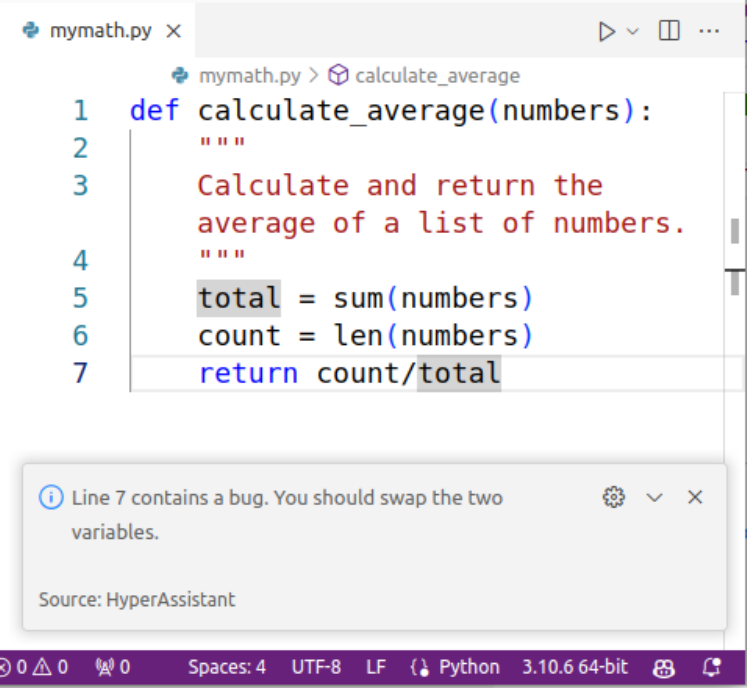}
		\caption{AI suggests a bug fix while the developer is coding.}
		\label{fig:bugfix}
	\end{subfigure}
	\par\bigskip 
	\begin{subfigure}[b]{.46\linewidth}
		\includegraphics[width=\linewidth]{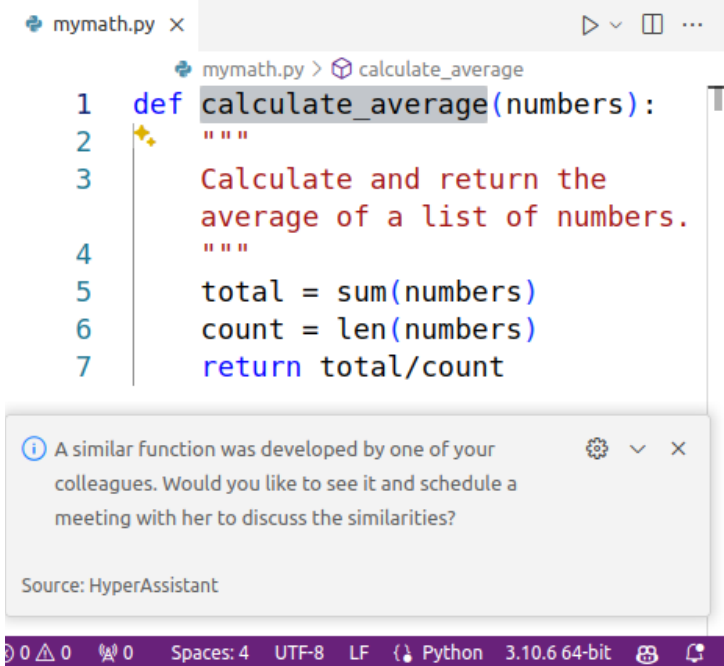}
		\caption{AI finds a similar method in the code base and suggests a team interaction.}
		\label{fig:similarintem}
	\end{subfigure}
	\hfill 
	\begin{subfigure}[b]{.45\linewidth}
		\includegraphics[width=\linewidth]{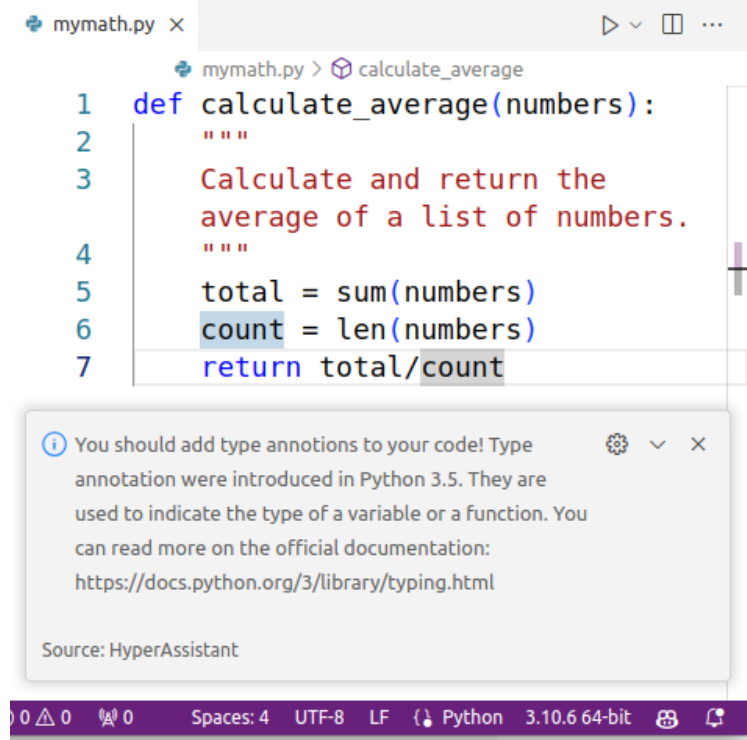}
		\caption{AI suggests a new feature with related documentation.}
		\label{fig:learning}
	\end{subfigure}
	\caption{Various \AI suggestions for improving developer working routine in 2030.}
	\label{fig:ai_suggestions}
\end{figure*}






\subsection{Mental Health}

The integration of \AI could emerge as a transformative ally for mental health, offering innovative approaches to support and prioritize developers' well-being, as it is an essential topic for their productivity~\cite{graziotin2017unhappy}.
Here, we delve into three critical areas where \AI interventions can significantly impact mental health and the general well-being of software developers.

First, developers often find themselves immersed in coding sessions for many hours, leading to mental fatigue and decreased productivity. \AI can monitor developers' activity levels and cognitive performance in real-time, identifying signs of fatigue or stress. For example, \AI algorithms can analyze typing patterns, code quality metrics, and even biometric data to detect when a developer might benefit from a break. By suggesting timely breaks, such as recommending a short walk or a brief mindfulness exercise, \AI helps developers rejuvenate their minds and maintain optimal focus throughout their workday. As a result, imagine a scenario where a developer has been coding for several hours and begins to make frequent syntax errors or experiences difficulty in concentrating. AI, equipped with machine learning models trained on developers' behavioral patterns, recognizes these signs of cognitive fatigue and prompts the developer to take a 10-minute break. During this break, AI suggests engaging in breathing exercises or listening to calming music, fostering relaxation and mental clarity upon return to work.

Second, \AI can enhance IDE environments by personalizing visual elements, such as color schemes and font sizes, to reduce eye strain and enhance readability. Additionally, AI-powered features can adjust ambient lighting and background music to create a more conducive atmosphere for concentration and positive emotions for developers~\cite{Girardi2022}. As a result, consider a developer who spends long hours coding late into the night. \AI, aware of the time and the developer's preferences, adjusts the IDE's color scheme to reduce blue light exposure, promoting better sleep quality. Furthermore, based on the developer's music preferences and mood indicators, \AI selects instrumental tracks with a soothing tempo to create a calming ambiance conducive to focused work.

Last, developers' mental and physical health are influenced by various factors beyond their coding activities, including exercise routines, dietary habits, and personal stressors. \AI can analyze developers' lifestyle data, such as fitness tracker metrics and self-reported wellness indicators, to offer personalized suggestions tailored to their unique needs and preferences. As a result, suppose a developer has been experiencing increased stress levels due to a combination of work pressure and personal commitments. \AI, integrated with the developer's calendar and fitness tracker, recognizes patterns indicating high stress levels and suggests incorporating short exercise breaks or mindfulness practices into their daily routine. Additionally, \AI may recommend healthy meal options or provide tips for improving sleep hygiene, addressing holistic aspects of the developer's well-being beyond the confines of coding tasks.

Through these examples, it becomes evident that AI-driven interventions have the potential to profoundly impact developers' mental health by providing proactive support, optimizing their work environment, and addressing broader lifestyle factors. Figure~\ref{fig:mentalhealth} shows an example of \AI for this task. By prioritizing mental well-being alongside technical proficiency, developers can foster a healthier and more sustainable approach to software development, ultimately leading to enhanced creativity, productivity, and job satisfaction.

\begin{solution}
    1~\AI has the potential to profoundly impact developers' mental health by providing proactive support and optimizing their work environment.
\end{solution}

\subsection{Fault Detection}
In the dynamic landscape of software development, the need for bug-free code remains essential~\cite{ketai2024}. \AI emerges as a transformative force, offering novel opportunities for bug detection and bug fixing. Here, we delve into critical tasks where \AI stands to revolutionize bug detection: vulnerability recognition, static analysis integration, and test case generation. Through these endeavors, \AI empowers developers to improve software quality and assist in delivering more resilient software solutions.

First, \AI can analyze the codebase for potential security vulnerabilities by recognizing patterns indicative of common security flaws. For instance, it can detect simple logic errors in real time as Figure~\ref{fig:bugfix} shows or even complex errors such as instances where user input is not properly sanitized before being used in SQL queries, potentially leading to SQL injection attacks. Upon detection, AI can suggest fixes or propose code modifications to mitigate these vulnerabilities, such as using parameterized queries instead of concatenating strings for SQL statements. Moreover, AI can integrate with static analysis tools like type checkers or linters to perform comprehensive code analysis. For example, in Python development, AI can automatically run type checkers like MyPy to identify type inconsistencies or potential type-related errors in the codebase~\cite{digrazia2022fse, Chow2024}. By flagging such issues, developers can ensure code robustness and maintainability.

Second, upon completion of a function or a class, \AI can automatically execute it within a sandbox environment to assess its functionality. By running various test cases, \AI can simulate different input scenarios to detect any unexpected behavior or failures. For instance, if a function is supposed to sort an array, \AI can generate random arrays of varying sizes and contents to validate the sorting algorithm's correctness. 

Last, \AI can assist in generating test cases automatically based on code coverage analysis and behavior prediction. By analyzing the code structure and potential execution paths, AI can generate test inputs that aim to maximize code coverage and uncover edge cases. This proactive approach to test case generation can help improve software quality by identifying bugs early in the development cycle.

\begin{solution}
    2~\AI will offer advanced capabilities for bug detection and bug fixing even for complex software systems.
\end{solution}

\subsection{Code Optimization}

Generating optimized code and reducing development times, for example reusing existing and efficient code, is crucial.
In 2030, \AI will be able to monitor in real-time the code written by developers, being able to find whether the same code is present in the code base, thus preventing its duplication. Figure~\ref{fig:similarintem} shows an example.
Moreover, \AI will also monitor online resources, like StackOverflow, and will suggest to the developer a specific snippet available online, dramatically reducing the development time.

This support can be extended also to related activities, like commenting and refactoring. Comments are a crucial component in software development, able to convey meaningful information and make the code clearer. However, sometimes they become obsolete, with developers that change the code without updating existing comments or javadoc~\cite{liu:tse2021}. This behavior is problematic and can have dangerous side effects, increasing the number of bugs in the code. 
\AI will help developers, ensuring consistency of comments with the written code, and raising warnings where issues are detected. \AI will also automatically update the comments, depending on the preference of each developer, even proposing a refactoring of the current solution and suggesting meaningful names for the variables.

\begin{solution}
    3~\AI will be able to fully support the developers for code optimization, favoring the usage of existing code and the alignment between code and comments.
\end{solution}

\subsection{Smart Team Interactions}

With the emergence of tools facilitating collaboration among numerous developers on the same project, like GitHub, ensuring an efficient interaction is essential.

\AI will boost this aspect, by favoring a more intense and synergical communication between developers, also improving the task assignment.
 \AI will be able to suggest a developer to ask the support of another team mate for writing a specific function, since it can realize that the code the developer is writing is more familiar or already written as Figure~\ref{fig:similarintem} shows.
\AI will be fully integrated in the team, monitoring the activities of all the members and promoting smart interaction between them. It will be able to estimate the development time required for a specific task, based on the long observation of her programming activities, optimizing the overall coding pipeline.

Moreover, \AI will be used by the team to generate a first draft of the entire project: starting from the abstract design, \AI can write the entire code of the project and then can assign each part to a specific developer, the one who is more suitable based on her background knowledge, to check its correctness.
\AI will also act like a reminder of certain tasks that are often forgotten by developers. It can, for example, monitor the activity of the developer and suggest to her that it is time to commit the results on GitHub, since it has completed the implementation of a specific function. It can also automatically schedule meetings between team members if it believes that this can be useful, in order to facilitate and improve the software development.

\begin{solution}
    4~\AI will improve the interaction between developers, leading to an improved task distribution and translating abstract ideas into code.
\end{solution}

\subsection{Learning and Developing New Skills}

Although programming languages evolve rapidly, developers' knowledge does not always keep the same pace, often remaining rooted in old yet reliable code.
Ciniselli \etal~\cite{ciniselli:arxiv2024} investigated the effect of the evolution of programming languages in the contest of AI tools, also showing that most of the methods extracted from GitHub belong to Java 8, released in 2014. This highlights an interesting trend for developers, that tend to stick to an old stable version rather than experimenting with novel features. 

\AI can help in filling this gap, suggesting interesting articles to the developer, related to the code she is writing, and even recommending a new feature that may be useful in that situation. \AI can even propose learning courses or tutorials, helping the developer to fill skill gaps in specific areas and generating ad-hoc tests for assessing the progress. 

\begin{solution}
    5~\AI will assist developers in honing their skills, proposing tailored learning resources promoting innovative features of the programming language.
\end{solution}

%% file: sections/2024vs2030.tex
\section{Developer Daily Routine: 2024 vs 2030}
\label{sec:2024vs2030}

In this section, we envision a hypothetical working day for a developer in 2024 and in 2030. This description focuses on the limitations of the AI assistants in 2024, showing all the potential benefits of the novel technologies that will be developed in the future. In our example, \AI can seem intrusive, but each developer can define the information it can access to, thus personalizing their coding and life experience.


\begin{table*}[ht]
    \centering
    \caption{Comparison between developer working routine in 2024 and 2030.}
    \label{tab:comparison}
    \begin{tabularx}{\textwidth}{lX X}
    \toprule
    {\bf AI Assistance} & {\bf AI Limitations in 2024} & {\bf AI Solutions in 2030} \\
    \midrule
    Mental Health & AI is not able to improve developers' mental health & AI can suggest the right moment for breaks and personalized activities to improve their well-being \\
    \midrule
    Fault Detection & AI is limited in automatically finding bugs and providing bug fixes for complex software systems & AI can automatically detect bugs and vulnerabilities, even recommending how to handle them \\
    \midrule
    Code Optimization & AI can recommend only simple code suggestions & AI can optimize the code, monitoring coding in real-time and suggesting alternatives \\
    \midrule
    Smart Team Interactions & AI is not able to handle or suggest interactions between colleagues or teams in the same company & AI is able to support developers by arranging useful meetings and fostering developers' interactions \\
    \midrule
    Learning New Skills & AI cannot suggest relevant resources or novel programming language features or APIs & AI can find alternatives involving new features, proposing tailored learning paths for developers \\
    \bottomrule
    \end{tabularx}
\end{table*}

\subsection{Developer Daily Routine in 2024}

Ashley, the developer in 2024, arrives in the office at 8a.m. in the morning. She immediately notices that the code she has written the day before has been changed by a colleague. Obviously, no comment at all! She spends 45 minutes trying to figure out the meaning of that code and finally she is ready to code. The task is quite demanding and she spends a couple of hours for the implementation of the new feature needed in this project. Clearly, a few typos in the code results in compilation errors, but Ashley patiently resolves them one by one. Finally it is compiling, but a few test cases give unexpected results. 
She spends several minutes asking GitHub Copilot to understand what was wrong with the code but the answers were too generic and useless so she decided to find the error by herself.

Stressed after 30 minutes of unsuccessful attempts at fixing that bug, she decides that it is time for a coffee break. In the coffee room, she meets Emma, a senior developer of her team, and she asks for help. Unfortunately, Emma is pretty busy so they can arrange a meeting for 2p.m, right after lunch. Ashley has lunch lost in her thoughts, trying to understand what was the mistake. The meeting with Emma is extremely helpful. After several minutes of intense concentration, they realize that a javadoc Ashley has taken by correcting is outdated and not aligned with the code so she has to fix the code. Emma also suggests looking at a new API released in the last version of Java that is faster in case she needs to speed up the computation. Unfortunately, Emma has no time for further help since she has a really busy agenda.
Ashley decides to start fixing the current version of the method since she is not aware of the new API and, after a stressful day, she does not want to spend time on a new learning task.

After one hour of effort, the code is working but, as Emma perceived, it is too slow. So she looks online to understand the  new suggested API, but the resources are limited and unclear, and it requires more time than expected. 
Finally, the working day is over and now the code seems to work properly, even though an unfixable warning message leaves her pondering. Ashley can finally go home after a stressful and non productive working day. And tomorrow (unfortunately) will be the same.

\subsection{Developer Daily Routine in 2030}

Ashley, the developer in 2030, arrives in the office and immediately notices that some code has changed since yesterday. However, thanks to \AI, a concise summary is presented to her, highlighting only the pertinent edits. With this efficiency, she swiftly comprehends the updates and is ready to begin her tasks.

As she starts coding, an intelligent bug detection system notifies her of an error she inadvertently introduced. The system not only reports the bug, but also suggests potential fixes, streamlining the debugging process. Furthermore, Ashley receives a notification regarding misalignment between the code and its corresponding javadoc comments. \AI offers suggestions on how to align them properly, ensuring code clarity and documentation consistency.

During her work, \AI recommends a piece of code implemented by a senior developer in the same company, recognizing its relevance to Ashley's task. \AI schedules a meeting for them by accessing their calendar, providing Ashley with preparatory materials to review beforehand. Additionally, Ashley is encouraged to take a break before the meeting, as \AI detects signs of fatigue, such as typos or syntax errors during the last 10 minutes, due to a lack of sleep the previous night accessing data from her wearable.

Following the productive meeting, Ashley successfully incorporates the optimized code, enhancing the project's features. With the task completed, AI suggests a balanced lunch from the Company Restaurant website, considering Ashley's plans for an evening gym session to maintain her well-being.

What once constituted a full working day for a developer in 2024 is now efficiently accomplished in half a day, allowing Ashley to tackle more tasks with precision, collaboration, and self-care in mind.

\section{Discussion and Future Work}
\label{sec:discussion}
The comparison between the working days of a developer in 2024 and 2030, summarized in Table \ref{tab:comparison}, sheds light on the remarkable advancements in AI technologies and their impact on the developer's productivity. As a result, we discuss implications for developers and researchers in the community of software engineering.


\subsection{Implications for Developers}

In 2024, Ashley's day is characterized by manual efforts and limited support from technology. She encounters challenges such as a huge list of code changes made by colleagues, debugging errors independently, and struggling with outdated documentation. While she seeks assistance from a senior developer, Emma, the help is constrained by Emma's busy schedule. Ashley's reliance on online resources for learning new APIs further slows down her progress. Despite her efforts, Ashley faces a stressful and unproductive day.

Conversely, in 2030, Ashley's experience is drastically different due to advancements in AI assistance. \AI streamlines her workflow by providing concise summaries of code changes and intelligent bug detection, significantly reducing the time spent on understanding modifications and debugging. The system also offers proactive suggestions for aligning code with documentation, enhancing code clarity and consistency. Furthermore, \AI facilitates collaboration by recommending relevant code implementations from within the company and scheduling meetings with colleagues, enabling efficient knowledge sharing and problem-solving. Moreover, \AI demonstrates a personalized approach by considering Ashley's well-being, detecting signs of fatigue, and recommending breaks and balanced meals. By leveraging data from wearable and company resources, \AI optimizes Ashley's productivity and supports her overall health. 

Overall, the comparison highlights the transformative impact of AI technologies on developer workflows in 2030. Developers like Ashley can accomplish tasks more efficiently, collaborate effectively, and prioritize self-care, leading to increased productivity and job satisfaction. The limitations and challenges faced by developers in 2024 underscore the significance of advancements in AI-driven assistance, illustrating the potential benefits of embracing novel technologies in the workplace.

\subsection{Implications for Researchers}

The comparison between developer working routines in 2024 and those projected for 2030, as detailed in Table \ref{tab:comparison}, not only underscores the transformative impact of AI on software engineering but also delineates crucial areas for future research.

Foremost, the evolution of AI in enhancing developers' mental health—from its initial inadequacy to a future where it intuitively recommends breaks and well-being activities—necessitates research into AI systems that can intricately understand and respond to individual health indicators. Practical exploration in this domain could involve collaborations between psychologists and software engineers to develop AI models that integrate psychological insights with real-time data analysis for personalized mental health recommendations.

The leap from basic fault detection to advanced, automated bug identification and resolution by 2030 invites the development of complex algorithms capable of deep code analysis. A practical research direction could involve partnerships with industry stakeholders to integrate these AI systems into existing development environments, enabling real-time, context-aware debugging suggestions based on historical project data and developer preferences.

In the sphere of code optimization, the shift toward AI that provides context-aware coding suggestions indicates a need for AI assistants that comprehend coding patterns, project intricacies, and optimization opportunities. Collaborative research with open-source communities could yield AI tools that learn from vast repositories of code to offer optimization advice, potentially even contributing code directly to projects.

The anticipated enhancement in smart team interactions through AI, transitioning from a non-existent to an active role in arranging and enhancing collaborations between developers, points to a future where AI aids in project management and team dynamics. Practical application of this research could see the creation of AI-driven platforms that analyze team performance data and project timelines to suggest optimal collaboration models and work distributions.

Lastly, the transition from AI's limited capability in suggesting new learning resources to a bespoke learning ecosystem tailored to developers' needs underscores the significance of AI in continuous professional development. This direction could be practically pursued by establishing partnerships with educational institutions and online learning platforms, utilizing AI to create dynamic, personalized learning pathways that adapt to the evolving technological landscape and individual learner goals.

These outlined paths pave the way for a future where AI not only boosts developer productivity but also plays a pivotal role in their professional and personal growth. Engaging in these research endeavors is crucial for unlocking AI's full potential in software engineering by 2030.

\subsection{Technical and Research Challenges}

The significant gap between the 2024 tools' capabilities and our vision for 2030 emphasizes some challenges that need to be addressed. Today's models lack in generating customized code that is adapted to a specific developer. They are not able to retain a significant amount of information about each developer, thus preventing the generation of suggestions that are familiar and adapted to their personal skills. This poses some challenges also for what concerns the improvement of the interactions between different developers, with models that may struggle in promoting synergical interactions based on the capabilities of each team member. To handle this challenge, research may focus on new and improved ways for creating contextual information that the model AI models can retain. The enhanced contextual knowledge can be leveraged to create a profile for each developer, thus enabling personalized suggestions and favoring team interactions.

The significant gap between the 2024 tools' capabilities and our vision for 2030 emphasizes some challenges that need to be addressed. Today's models lack in generating customized code that is adapted to a specific developer. They are not able to retain a significant amount of information about each developer, thus preventing the generation of suggestions that are familiar and adapted to their personal skills. This poses some challenges also for what concerns the improvement of the interactions between different developers, with models that may struggle in promoting synergical interactions based on the capabilities of each team member. To handle this challenge, research may focus on new and improved ways for creating contextual information that the model AI models can retain. The enhanced contextual knowledge can be leveraged to create a profile for each developer, thus enabling personalized suggestions and favoring team interactions.

Another limitation lies in the optimization of the generated code. AI assistants learn from the code used during the training, that can sometimes be low-quality, containing bugs or vulnerabilities. Hence, they also tend to recommend sub-optimal suggestions that can introduce serious vulnerability issues. A viable solution to mitigate this problem is the improvement of models' reasoning abilities, allowing them to infer possible drawbacks of the proposed solution and explore the effectiveness of alternative approaches.

%% file: sections/conclusion.tex
\section{Conclusion}
\label{sec:conclusion}
In this paper, we discuss how AI for Software Engineering can bridge the gap between the current limitations and the future potential in areas such as mental health support, fault detection, code optimization, team interaction, and learning new skills. The transition from a reactive to a proactive AI approach in software engineering underscores the technology's capability to not only understand and adapt to the individual needs of developers but also to foster a collaborative, efficient, and health-conscious working environment.

Furthermore, we discuss how AI can serve as a catalyst for interdisciplinary research, merging insights from psychology, education, and project management to create a holistic support system for developers. This collaborative approach is crucial for developing AI systems that are not only technically proficient but also attuned to the human aspects of software development.

In sum, as we discuss the capabilities of AI, it is clear that its integration into software development heralds a new era of efficiency and well-being. The continued exploration of AI's potential will undoubtedly lead to significant advancements, making the profession more fulfilling and productive.